\documentclass[prd,superscriptaddress,nofootinbib,preprintnumbers,amssymb]{revtex4}
\usepackage{latexsym}
\usepackage{dcolumn}
\usepackage{epsfig}

\makeatletter
\def\vereq#1#2{
\lower3pt\vbox{\baselineskip1.5pt \lineskip1.5pt
\ialign{$\m@th#1\hfill##\hfil$\crcr#2\crcr\sim\crcr}}}
\makeatother

\begin{document}

\preprint{FERMILAB-Pub-02/298-T}
\preprint{{UMD-PP-03-027}}

\title{Manifest CP Violation from Majorana Phases}

\author{Andr\'e de Gouv\^ea}
\affiliation{Theoretical Physics Department, Fermilab, 
P.O. Box 500, Batavia, IL, 60510-0500, USA }

\author{Boris Kayser}
\affiliation{Theoretical Physics Department, Fermilab, 
P.O. Box 500, Batavia, IL, 60510-0500, USA }

\author{Rabindra N. Mohapatra}
\affiliation{ Department of Physics, University of Maryland, College
Park, MD-20742, USA}

\begin{abstract}
We hunt for and discuss manifestly CP-violating effects which are
mediated by Majorana phases. These phases are present if the Standard Model
neutrinos are Majorana particles. We argue that while Majorana phases do
affect the strength of neutrinoless double beta decay (a well known fact), 
they do so in a way that involves no manifest violation of CP. 
The conditions for manifestly CP-violating phenomena -- differences between the rates
for CP-mirror-image processes -- are presented, and three examples are discussed:
(i)~neutrino~$\leftrightarrow$~antineutrino oscillation; (ii)~rare
decays of $K$ and $B$ mesons and their antiparticles and (iii)~the lepton asymmetry
generated by the decay of hypothetical very heavy right-handed ``see-saw'' neutrinos. 
We also find that, for the case of degenerate light neutrinos, manifestly CP-violating 
effects in neutrino~$\leftrightarrow$~antineutrino oscillation vanish, although flavor-changing
transitions do not. Finally, we comment on leptogenesis with degenerate right-handed neutrinos,
and contrast it to the neutrino~$\leftrightarrow$~antineutrino oscillation case. 
\end{abstract}

\maketitle

\setcounter{equation}{0}
\section{Introduction}

If neutrinos are Majorana particles, then the leptonic mixing matrix $U$ can 
contain more CP-violating phases than its quark counterpart (for the same
number of generations) \cite{phases}. The additional phases, known as Majorana
phases, have no effect on neutrino oscillation. Indeed, the only current or
proposed neutrino experiment that could in principle provide evidence of Majorana phases is
the search for neutrinoless double beta decay, $0\nu\beta\beta$ \cite{betabeta_exp}. 
The rate for this process depends not only on the neutrino masses and mixing angles, but also 
on CP-violating phases, notably including the Majorana phases. However, even if
experimental and theoretical uncertainties should permit us to obtain evidence for a
non-vanishing Majorana phase from the rate of $0\nu\beta\beta$ \cite{0nubb}, the effect of 
CP-violating phases on this reaction is not a manifestly CP-violating phenomenon. 
By the latter, we mean a CP-odd effect -- 
a difference between the rate for some physical process and that for its CP-mirror
image. While CP-odd phases in the leptonic mixing matrix do affect the rate $\Gamma$ for
$0\nu\beta\beta$, they do so in a CP-even way; that is, their effect on the rate $\Gamma$
for some particular nuclear double beta decay is the same as on the rate $\bar{\Gamma}$
for the CP-mirror-image decay (the decay of an antinucleus), so that $\Gamma=\bar{\Gamma}$.
Therefore, even if we could study the neutrinoless double beta decay of antinuclei (an
impossibility in practice, to say the least), we would be unable to observe a ``smoking
gun'' signal of CP-violation due to Majorana phases. 

In this paper, we ask whether Majorana phases, like the more familiar CP-violating
``Dirac'' phase in the quark mixing matrix, can lead to CP-odd effects. If so, where and
under what conditions can these effects occur, and what are they? 
Are they observable in practice?

An increasingly appealing explanation of the present baryon asymmetry in the Universe 
rests on early-universe ``leptogenesis,'' resulting from CP violation in the 
decays of so-far hypothetical, very heavy Majorana neutral leptons \cite{leptogenesis}.
The required CP violation in this process can come from Majorana phases. Furthermore, it
is a CP-odd effect -- a difference between two CP-mirror-image decays, one of which
yields a lepton, the other an antilepton. Thus, Majorana phases can, in principle,
yield CP-odd effects. However, the Majorana phases that act in the early Universe
are not those in the mixing matrix $U$ that governs light neutrino mixing 
\cite{heavy_light}. Moreover, the role of these ``early-universe phases'' depends
on the existence of hypothetical heavy Majorana leptons.\footnote{The existence of
heavy ``right-handed neutrinos'' is strongly motivated by the see-saw mechanism for 
generating light neutrino masses \cite{seesaw}. Unfortunately, even if this beautiful 
theoretical idea is correct, we may never be able to observe direct evidence for the 
existence of heavy right-handed neutrinos if their masses are indeed many orders of
magnitude above the weak-scale, as naively indicated by the experimental evidence for
neutrino masses.} We therefore ask whether the Majorana phases in the light-neutrino
mixing matrix $U$ can lead to CP-odd effects that depend only on the (assumed) Majorana
nature of the light neutrinos, and not on the existence of any additional Majorana 
particles. 

We find that the answer is yes -- Majorana phases in $U$ can induce CP-odd effects. In 
particular, they do so in the process of ``neutrino~$\leftrightarrow$~antineutrino 
oscillations'' \cite{neutrino_antineutrino}. By that we mean a process in which, for example, 
a neutrino ``beam'' is created by
incoming positively-charged leptons, but is measured in a detector via the production
of negatively-charged leptons. If Majorana phases are present, the rates for this
process and for its CP-mirror image (where the charges of the charged
leptons are reversed) will, in general, differ. We explicitly point out why CP-odd
effects can occur in neutrino~$\leftrightarrow$~antineutrino oscillations but not 
in $0\nu\beta\beta$. We further discuss under what conditions the            
neutrino~$\leftrightarrow$~antineutrino oscillation process occurs, and when CP-odd
effects can be observed. For example, we point out that when 
the neutrino masses are degenerate, CP-odd effects disappear, but the 
neutrino~$\leftrightarrow$~antineutrino 
oscillation process can still take place if the Majorana phases are nontrivial. 
The rate for this process still depends on the mixing angles in the mixing matrix.
Thus, when Majorana phases are present, mixing angles continue to have physical
consequences even when the neutrino masses are degenerate. This 
simple yet remarkable behavior is in marked contrast to the behavior of quark mixing,
where flavor mixing would disappear (and the ``mixing angles'' become unphysical) if
all the charge-2/3 quarks or all the charge-(--1/3) ones were degenerate in mass. 

Other processes where leptonic CP-odd effects could appear are the rare, lepton-number 
violating
meson decays $K^\pm\rightarrow \pi^\mp\mu^\pm\mu^\pm $ and
$B^\pm\rightarrow \pi^\mp\tau^\pm\tau^\pm$. Here, the rate for the $K^+$ decay could differ 
from
that for the $K^-$ decay, and similarly for the $B^+$ and $B^-$ decays. However, we find that
new sources of lepton-number violation (on top of the neutrino Majorana masses) are
required in this case.

By picturing the neutrino~$\leftrightarrow$~antineutrino oscillation process in terms
of Feynman diagrams, and rearranging the pieces of these diagrams, 
we show that the CP-odd effect resulting in leptogenesis grows
out of the Majorana phases in exactly the same way as the CP-odd effect in  
neutrino~$\leftrightarrow$~antineutrino oscillation. This leads us to investigate whether
leptogenesis also ``disappears'' when the masses of all the heavy Majorana leptons 
are of equal magnitude. We discuss under which conditions this would indeed be the case.

Unfortunately, if neutrinos interact only via Standard Model left-handed interactions, 
neutrino~$\leftrightarrow$~antineutrino oscillations, while yielding interesting 
conceptual insights into the possible effects of Majorana phases, are virtually 
unobservable in practice. This is due to the fact that, because we consider the
neutrino Majorana masses to be the source of lepton-number violation,
the rate for neutrino~$\leftrightarrow$~antineutrino oscillations is suppressed by powers of 
the neutrino masses (in units of the neutrino total energy).  
This is also true of the rates for the $\Delta L=2$ decays 
$K^\pm\rightarrow\pi^\mp\mu^\pm\mu^\pm $ and
$B^\pm\rightarrow \pi^\mp\tau^\pm\tau^\pm$. Since these rates are 
proportional to positive powers of the neutrino masses, their associated 
branching ratios are expected to be of $O(10^{-22})$ (the present
upper-limits on these branching ratios are $O(10^{-9})$ \cite{PDG}).
However, even though all these processes are unlikely to be observable in the foreseeable 
future, they provide clear illustrations of how, in principle, Majorana phases can lead to 
manifestly CP-violating effects in low-energy reactions.

Our presentation is the following: First, we define Majorana phases and discuss
when they are potentially observable. Second, we discuss in some detail the process of
neutrinoless double beta decay, and explain why CP-odd effects would not be present even if
antinuclear double beta decay could be observed. We then proceed to outline the requirements
for observing manifest CP-odd effects, and discuss in detail 
neutrino~$\leftrightarrow$~antineutrino oscillations. 
Next, we show how CP phases in the neutrino sector can manifest themselves in differences 
between the rates for the
lepton-number violating decay rates of $K^+$ and $K^-$ and $B^+$ and $B^-$,
and under what conditions.   
Finally, we comment on the relation between neutrino~$\leftrightarrow$~antineutrino 
oscillations and
heavy right-handed Majorana neutrino decays, paying special attention to the 
dependency of both processes on Majorana phases and the effect of mass-degenerate 
neutrino states.  

\setcounter{equation}{0}
\section{Majorana Phases}
\label{sec_maj}

We assume that each neutrino mass eigenstate $\nu_i$, $i=1,2,3\ldots$ (with mass $m_i$), 
is a Majorana fermion. This means that $\nu_i$ is its own antiparticle, so that its 
field is its own charge conjugate, up to a phase factor. Thus,
\begin{equation}
\nu_i=\lambda_i\nu_i^c,
\label{phase_factor}
\end{equation}
where the superscript $c$ denotes charge conjugation and $\lambda_i$ is a phase factor
henceforth referred to as the charge conjugation phase factor. 

We further assume that the neutrino coupling to charged leptons and the $W$-boson is
as prescribed by the Standard Model (SM). For massive neutrinos, the SM interaction
Lagrangian is 
\begin{equation}
{\cal L}_{\rm int}=-\frac{g}{\sqrt{2}} W_{\rho}^-\sum_{\alpha,i}\bar{l}_{\alpha L}\gamma^{\rho}
U_{\alpha i} \nu_{i L} -\frac{g}{\sqrt{2}} W_{\rho}^+
\sum_{\alpha,i}\bar{\nu}_{i L}U_{\alpha i}^*\gamma^{\rho} l_{\alpha L}.
\label{inter}
\end{equation}
Here, $g$ is the semiweak coupling constant and $\alpha$ runs over the charged lepton flavors:
$\alpha=e,\mu,\tau\ldots$ The subscript $L$ denotes chiral left-handed projection, and
$U$ is the unitary leptonic mixing matrix. It will become useful later to rewrite
\begin{equation}
\bar{l}_{\alpha L}\gamma^{\rho}U_{\alpha i} \nu_{i L}=-\bar{\nu^c}_{i R}\gamma^{\rho}
U_{\alpha i}l^c_{\alpha R}=-\lambda_i\bar{\nu}_{i R}\gamma^{\rho}
U_{\alpha i}l^c_{\alpha R},
\label{inter_R}
\end{equation}
where $\lambda_i$ is the charge conjugation phase factor defined in Eq.~(\ref{phase_factor})
and $R$ denotes right-handed chiral projection. 

As is the case in the quark sector, the leptonic mixing matrix is written in the
basis where the charged lepton and (Majorana) neutrino mass matrices are 
real, positive, and diagonal. Generically, it can be written in the form
\begin{equation}
U=\mathbb{E}^{i\frac{\phi_{\alpha}}{2}}U'\mathbb{E}^{i\frac{\xi_{i}}{2}},
\label{extract}
\end{equation}
where $\mathbb{E}^{i\phi_{\alpha}/2}={\rm diagonal}(e^{i\phi_e/2},e^{i\phi_{\mu}/2},
\ldots)$, $\mathbb{E}^{i\xi_{i}/2}={\rm diagonal}(e^{i\xi_1/2},e^{i\xi_2/2},\ldots)$ 
are diagonal ``phase matrices,'' and $U'$ is a (non-generic) unitary mixing matrix. 
Within the SM, 
the phases contained in $\mathbb{E}^{i\phi_{\alpha}/2}$ are not physical, as they can
be ``absorbed'' by redefining the right-handed charged lepton fields (which do not
feel the charged-current weak interactions). The phases $\xi_i$ are potentially observable,
and will henceforth be {\sl defined as ``Majorana phases.''} For example, in the case
of two lepton species, $U'$ is real and parametrized by one mixing angle, 
while there is one potentially observable Majorana phase $\xi\equiv\xi_{2}-\xi_1$.\footnote{
An overall phase, common to all neutrinos, is not physical. One is only sensitive
to phase differences.} It is
important to stress that if the neutrinos were Dirac particles, all phases contained
in $\mathbb{E}^{i\xi_{i}/2}$ would also be unphysical, as they could be ``absorbed''
by redefining the SM singlet right-handed neutrino fields.

A Majorana phase is therefore characterized as one that is common to all elements
of a given column of the leptonic mixing matrix, as defined in Eq.~(\ref{inter}). That is,
it is a phase that affects all $U_{\alpha i}$ equally, irrespective of the flavor $\alpha$,
for a given neutrino mass eigenstate $\nu_i$. Of course, elements of a column 
of $U$ may contain additional phases that are not common to the whole column (these are
the ``left-over'' phases contained in $U'$, as defined in Eq.~(\ref{extract})). These 
phases will be {\sl defined as ``Dirac Phases.''}\footnote{The reason for the definition
should be clear. If the neutrinos were Dirac fermions, 
all would-be Majorana phases could be ``absorbed'' 
by appropriately redefining the neutrino fields, and the only observable CP-odd effects 
would be parametrized by the Dirac phases.} 

As is well known, if $U$ is not real ({\it i.e.}, it contains nontrivial phases), 
the physical processes mediated by Eq.~(\ref{inter}) need not be 
CP-preserving.\footnote{See \cite{CP_book} for a pedagogical discussion of the conditions 
which 
the massive neutrino Lagrangian must satisfy in order to necessarily conserve CP.} However,
while Dirac phases can lead to CP-noninvariance irrespective of the nature of the neutrinos,
Majorana phases can do so only if the neutrinos are Majorana particles. It is interesting
to understand the origin of this fact by comparing a process that can occur regardless 
of the character of the neutrinos with a related one that can occur only if the neutrinos
are Majorana particles. 

\begin{figure}[t]
\vspace{1.0cm}
\centerline{\epsfxsize 7cm \epsffile{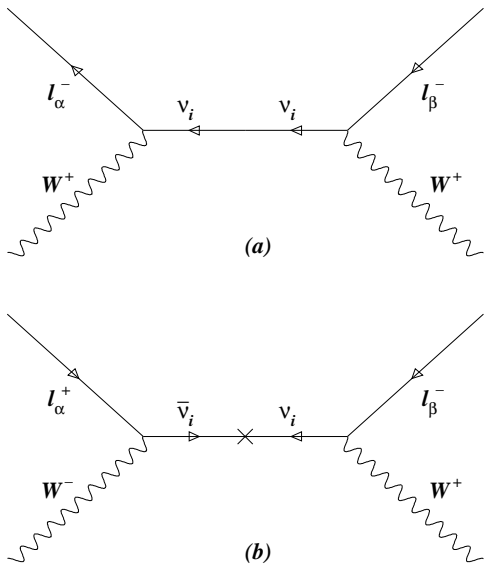}}
\caption{Feynman diagrams for $(a)$ neutrino~$\leftrightarrow$~neutrino oscillation
and $(b)$ neutrino~$\leftrightarrow$~antineutrino oscillation. Time flows from 
the left to the right, and the arrows represent
the chirality of the various fermions. The $\times$ indicates a chirality-flip in the
neutrino propagator, which is proportional to the neutrino Majorana mass. Note that 
$\nu_i=\bar{\nu}_i$ up to a phase factor.}
\label{diagrams}
\end{figure}

First, we will analyse the process of ``neutrino~$\leftrightarrow$~neutrino'' 
oscillation, for which there is ever-increasing experimental evidence. This oscillation
may be viewed as the process
\begin{equation}
l^-_{\alpha}W^+\rightarrow \nu \rightarrow l_{\beta}^-W^+,
\label{osc}
\end{equation}
in which the intermediate-state neutrino propagates a macroscopic distance $L$. This process
is depicted in Fig.~\ref{diagrams}a. 
The intermediate-state neutrino can be in any of the mass eigenstates $\nu_i$. Thus,
the amplitude $A_L$ for this lepton-number ($L$) 
conserving process may be written schematically as 
\begin{equation}
A_L=\sum_i \langle l_{\beta}^- W^+ | H_{\rm int} 
| \nu_i\rangle \langle \nu_i | H_{\rm int} | l_{\alpha}^- W^+
\rangle, 
\label{A_L}
\end{equation}
where $H_{\rm int}$ is the interaction Hamiltonian associated with Eq.~(\ref{inter}).
Now, $\langle l_{\beta}^- W^+ | H_{\rm int} 
| \nu_i\rangle \langle \nu_i | H_{\rm int} | l_{\alpha}^- W^+
\rangle\propto U_{\beta i}U_{\alpha i}^*$ (see Eq.~(\ref{inter})). Thus, even if the
$i$th column of $U$ contains the Majorana phase factor $e^{i\xi_i/2}$, it is clear that 
it will cancel out of $U_{\beta i}U_{\alpha i}^*$ ($\forall\beta,\alpha$) and will 
consequently have no effect on the amplitude $A_L$.

Next we analyse a qualitatively different process: 
``neutrino~$\leftrightarrow$~antineutrino'' oscillation \cite{neutrino_antineutrino}. 
This is the reaction
\begin{equation}
l^+_{\alpha} W^-\rightarrow \nu \rightarrow l_{\beta}^- W^+,
\label{anti_osc}
\end{equation}
in which, once more, the intermediate neutrino travels a macroscopic distance
$L$. This process is depicted in Fig.~\ref{diagrams}b.  

Unlike ordinary flavor oscillations (Eq.~(\ref{osc})), the process 
Eq.~(\ref{anti_osc}) can only occur if lepton number is
no longer a good quantum number. This is exactly the case if the neutrinos have 
non-vanishing Majorana masses, which also implies that the neutrino mass eigenstates are
Majorana particles.\footnote{If CPT is also broken, the mass eigenstates are {\sl not} 
Majorana particles even in the presence of Majorana mass terms \cite{BBBK}. We will assume
throughout this paper that CPT is a good symmetry.} 
As in ordinary flavor oscillations, the intermediate neutrino in Eq.~(\ref{anti_osc}) can be 
in any of the mass eigenstates $\nu_i$. Thus, the amplitude $A_{\not{L}}$ for the
lepton-number violating process Eq.~(\ref{anti_osc}) can be written 
schematically
\begin{equation}
A_{\not{L}}=\sum_i \langle l_{\beta}^- W^+ | H_{\rm int} 
| \nu_i\rangle \langle \nu_i | H_{\rm int} | l_{\alpha}^+ W^-
\rangle.
\label{A_nonL}
\end{equation}
As before, 
$\langle l_{\beta}^- W^+ | H_{\rm int} | \nu_i\rangle\propto U_{\beta i}$. However, the second
bracket in Eq.~(\ref{A_nonL}) requires some care. Like the first bracket,
$\langle \nu_i | H_{\rm int} | l_{\alpha}^+ W^- \rangle$ also comes from the first
term in Eq.~(\ref{inter}), but one should use the term as it was rewritten in 
Eq.~(\ref{inter_R}). This is so that the field $\bar{\nu}_i$ in the second term of 
Eq.~(\ref{A_nonL}) can be contracted with the field $\nu_i$ in the first term in order
to make the usual neutrino propagator, $\langle 0|T(\nu_i\bar{\nu}_i)|0\rangle$. 
From Eq.~(\ref{inter_R}), $\langle \nu_i | H_{\rm int} | l_{\alpha}^+ W^- \rangle\propto
\lambda_i U_{\alpha i}$, so that
\begin{equation}
A_{\not{L}}=\sum_i \left( \lambda_i U_{\alpha i} U_{\beta i} \right) K_i,
\label{A_interf}
\end{equation}  
where $K_i$ is a kinematical factor.

The phases of the $U_{\alpha i}$ and of $\lambda_i$ all depend on the phase convention
chosen for the state $|\nu_i\rangle$. However, it is readily shown that the product 
$\lambda_i U_{\alpha i} U_{\beta i}$ is phase-convention-free 
\cite{convention}.\footnote{Redefining the Majorana phases by rotating the state
$|\nu_i\rangle$ so that $U_{\alpha i}\rightarrow U_{\alpha i}e^{i\theta_i}$ also leads to 
$\lambda_i\rightarrow \lambda_i e^{-2i\theta_i}$.} This means that the interference
of the different terms that contribute to $A_{\not{L}}$ in Eq.~(\ref{A_interf}) can
lead to convention-free physical effects, which clearly depend on the Majorana phases 
of $U$. A Majorana phase factor in the $i$th column of $U$ should be thought of as
the phase factor present in this column for a fixed value of $\lambda_i$ corresponding
to the chosen phase convention for $|\nu_i\rangle$, By phase redefining $|\nu_i\rangle$,
we can always remove the Majorana phase from the $i$th column of $U$, but this phase
would then simply reappear in $\lambda_i$, leaving $A_{\not{L}}$ unchanged.

We conclude that when neutrinos are Majorana particles, the rates for 
lepton-number violating processes depend on the Majorana phases. 
Processes that do not involve lepton number violation in some form 
are not, at least at leading order, 
capable of exploring the leptonic mixing matrix in a way that would reveal
the presence of Majorana phases. It should be emphasized, as pointed out
by the authors of \cite{Nieves_Pal}, that the rates for lepton-number conserving
processes can show such a presence (which may lead to CP-odd effects \cite{NP_new}), 
provided that they receive
a significant contribution from processes that violate lepton number ``$1+(-1)$~times.''
These contributions, however, are very suppressed and unobservable under
most circumstances.

\setcounter{equation}{0}
\section{Neutrinoless Double Beta Decay}
\label{sec_doub}

The most promising way of probing the Majorana nature of the neutrino is to look
for neutrinoless double beta decay ($0\nu\beta\beta$). This is the lepton-number 
violating nuclear decay process $Z\rightarrow (Z+2)+e^-e^-$, where $Z~(Z+2)$ is the 
atomic number of the parent (daughter) nucleus. Assuming CPT-invariance, 
the observation of this process would demonstrate that neutrinos are indeed Majorana
particles \cite{Schechter_Valle}. If $0\nu\beta\beta$ does occur, it is very likely 
dominated by a mechanism in which the parent nucleus emits a pair of $W^-$ bosons,
turning into the daughter nucleus, and then the $W^-$ bosons exchange one or another neutrino
mass eigenstate to produce the two outgoing electrons.
The heart of the mechanism is the second step: $W^-W^-\rightarrow e^-e^-$ via
Majorana neutrino exchange. In the cross channel, this step is simply
$e^+W^-\rightarrow \nu \rightarrow e^-W^+$, {\it i.e.}, Eq.~(\ref{anti_osc}) for
$l_{\alpha}=l_{\beta}=l_e$. Of course, the neutrino now only
propagates a very short distance (of the size of a nucleus). 

Assuming that Majorana neutrino exchange is indeed the dominant contribution to 
$0\nu\beta\beta$, its amplitude $A_{\beta\beta}$ 
should be of the form Eq.~(\ref{A_interf}), and, indeed,
it is. As is well known \cite{BK84}, when neutrino exchange dominates,
\begin{equation}
A_{\beta\beta}=\sum_i \left(\lambda_iU_{ei}^2\right)m_i K,
\label{Abb}
\end{equation}
where $m_i$ is the mass of the neutrino mass-eigenstate $\nu_i$, which is chosen
to be real and positive, while $K$ is a kinematical and nuclear factor that does
not depend on $i$. The quantity
\begin{equation}
\left|\sum_i \left(\lambda_iU_{ei}^2\right)m_i\right|\equiv m_{\beta\beta}~(\equiv |M_{ee}|)
\label{Mbb}
\end{equation}
is known as the effective neutrino mass for neutrinoless double-beta decay, and is simply
the absolute value of the $ee$-element of the Majorana neutrino mass matrix in the
basis where the charged lepton mass-matrix and the $W$-boson couplings are diagonal.

Suppose, for the purpose of illustration, that there are three neutrino species, so 
that $U$ is a $3\times3$-matrix. Choose the phase convention where $\forall i,~\lambda_i=1$,
and take
\begin{equation}
U=U'\times{\rm diagonal}(e^{i\frac{\xi_1}{2}},e^{i\frac{\xi_2}{2}},e^{i\frac{\xi_3}{2}}),
\end{equation}
as in Eq.~(\ref{extract}). It is obvious from Eq.~(\ref{Abb}) that the overall
rate for $0\nu\beta\beta$, $\Gamma_{0\nu\beta\beta}=|A_{\beta\beta}|^2$, is affected
by (some combination of) the Majorana phases $\xi_i$. Therefore it is clear (and well
known) that Majorana phases lead to physical consequences.

It is interesting to notice that if $U$ contains a Dirac phase -- a CP-violating
phase that is not common to an entire column of $U$ -- as is generically
the case if there are at least three neutrino species, then this phase may also influence
$0\nu\beta\beta$ in the {\sl same} way that Majorana phases can. The amplitude for
$A_{\beta\beta}$ depends on the leptonic mixing matrix through $\lambda_iU_{ei}^2$, and
it makes no difference whether some phase factor appears in the entire $i$th column
of $U$ or only in $U_{ei}$. To be sure, if some $U_{ei}$ is proportional to a Dirac phase
factor $e^{-i\delta}$, one can always remove this factor through the $\nu_i$ phase 
redefinition $|\nu_i\rangle\rightarrow |e^{i\delta}\nu_i\rangle$. However, this
redefinition also results in $\lambda_i\rightarrow e^{-2i\delta}\lambda_i$, leaving
the phase convention-independent combination $\lambda_iU_{ei}^2$ unchanged.

Imagine now that one could measure the rate for the CP-mirror image of the decay
$Z\rightarrow (Z+2)+e^-e^-$, namely $\overline{Z}\rightarrow\overline{(Z+2)}
+e^+e^+$. In contrast to 
the $e^-e^-$ decay, this $e^+e^+$ decay involves the second term in Eq.~(\ref{inter}) and
the analog of Eq.~(\ref{inter_R}) for this term. Thus, in the amplitude $A_{\beta\beta}$
the factor $\lambda_iU_{ei}^2$ is replaced by $(\lambda_iU_{ei}^2)^*$ so that the amplitude 
$\bar{A}_{\beta\beta}$
for $\overline{Z}\rightarrow\overline{(Z+2)}+e^+e^+$ is
\begin{equation}
\bar{A}_{\beta\beta}=\sum_i\left(\lambda_iU_{ei}^2\right)^* m_i \bar{K}.
\label{anti_Abb}
\end{equation}
Here, due to the CP-invariance of the strong interactions that determine nuclear matrix 
elements, the kinematical and nuclear factor $\bar{K}$ is identical to $K$ in 
Eq.~(\ref{Abb}), except for a possible (irrelevant) phase difference. Thus, the rate for
``anti-$0\nu\beta\beta$'' is identical to the $0\nu\beta\beta$ rate:
$|A_{\beta\beta}|^2=|\bar{A}_{\beta\beta}|^2$. That is, while the Majorana phases
and, for that matter, the Dirac phases, affect the rate for $0\nu\beta\beta$,
they do so in a CP-even way: their effects on a given $0\nu\beta\beta$ process
and its CP-mirror-image anti-$0\nu\beta\beta$ process
are identical. The only way to determine the effects
of CP phases in neutrinoless double beta decay is to determine, through other 
experiments, the masses $m_i$ and the ``mixing angles'' $|U_{ei}|^2$ of the leptonic 
mixing matrix, and compare the results with the obtained measurement of 
Eq.~(\ref{Mbb}).\footnote{For example, if $m_{\beta\beta}\neq 
\sum m_i|U_{ei}|^2$, we could conclude 
unambiguously that some of the CP-odd phases $\xi_{ij}\equiv \xi_i-\xi_j$, $i,j=1,2,\ldots$ 
are different from $0~{\rm mod}~2\pi$.} Whether this can be accomplished in practice has 
recently been explored by several authors \cite{0nubb}.     

\setcounter{equation}{0}
\setcounter{footnote}{0}
\section{Manifest CP Violation from Majorana Phases in ``Low-Energy'' Phenomena}

A Dirac phase in the quark or lepton mixing matrix can certainly produce CP-odd effects. Can
a Majorana phase in the lepton mixing matrix do this too? In the rate for $0\nu\beta\beta$,
Majorana phases lead only to a CP-even effect, as we have just seen.

To try to find a process in which the effects of Majorana phases can include CP-odd ones,
we begin by asking what it takes to produce a CP-odd effect. If CP violation comes from 
phases, we can answer this question in a very general, well known way \cite{CP_book}. 
Suppose that some physical process
$P$ has an amplitude $A$ consisting of two contributions:
\begin{equation}
A=a_1e^{i\delta_1}e^{i\phi_1}+a_2e^{i\delta_2}e^{i\phi_2},
\label{P}
\end{equation} 
where $a_{1,2}$ are the magnitudes of the two contributions, $\delta_{1,2}$ are CP-odd 
phases which change sign when one computes the amplitude for the anti-process $\bar{P}$, 
while $\phi_{1,2}$ are CP-even phases that are the same for both $P$ and $\bar{P}$. The 
amplitude for $\bar{P}$ is, therefore,
\begin{equation}
\bar{A}=a_1e^{-i\delta_1}e^{i\phi_1}+a_2e^{-i\delta_2}e^{i\phi_2}.
\label{Pbar}
\end{equation} 
Note that the magnitudes $a_{1,2}$ are the same for $P$ and $\bar{P}$ because we are 
assuming that CP-violating effects come from phases.

The CP-odd difference $\Delta_{CP}\equiv|\bar{A}|^2-|A|^2$ is
\begin{equation}
\Delta_{CP}=4a_1a_2\sin\left(\delta_1-\delta_2\right)\sin\left(\phi_1-\phi_2\right),
\label{Delta_gen}
\end{equation}
where we used Eqs.(\ref{P},~\ref{Pbar}). It is clear that in order for there to be a CP-odd
effect in the rates for a process $P$ and its mirror-image, the amplitude for the process
must satisfy the following three requirements:
\begin{itemize}
\item{It must contain at least two distinct contributions.}
\item{The distinct contributions must be proportional to CP-odd phase factors with 
corresponding phases $\delta_i$, satisfying the condition(s) $\delta_i-\delta_j\neq 
0~{\rm mod}~\pi$ for some $i,j$.}
\item{These contributions must also be proportional to CP-even phase factors with 
corresponding phases $\phi_i$, satisfying the condition(s) $\phi_i-\phi_j\neq 
0~{\rm mod}~\pi$ for some $i,j$.}
\end{itemize} 

In leptonic processes such as $0\nu\beta\beta$, the necessary CP-odd phases may be provided
by the leptonic mixing matrix. However, from Eq.~(\ref{Abb}), it is easy to see that
the third of the requirements listed above is not satisfied. 
Note that the CP-even phases in Eq.~(\ref{Abb}) are $\phi_i={\rm 
arg}(K),~\forall i$. It is, therefore, 
clear that there are no CP-even phase differences. This is why phases in $U$
do not lead to CP-odd effects in $0\nu\beta\beta$.

\subsection{Neutrino~$\leftrightarrow$~Antineutrino Oscillations}
\label{sec_osc}

It should be clear that in order to observe CP-odd effects due to the Majorana phases in 
some process, the amplitude for that process must also contain CP-even phases which 
differ from one piece of the amplitude to another. With this requirement in mind, we turn 
to neutrino~$\leftrightarrow$~antineutrino oscillations, Eq.~(\ref{anti_osc}).

Assuming that the neutrinos travel a macroscopic distance, we would like to compute the
amplitude for the process Eq.~(\ref{anti_osc}), which is depicted in 
Fig.~\ref{diagrams}b. As discussed
in Sec.~\ref{sec_maj}, the amplitude $A_{\not{L}}$ has the general form 
Eq.~(\ref{A_interf})
and can be written as \cite{neutrino_antineutrino}
\begin{equation}
A_{\not{L}}=\sum_i \left(\lambda_iU_{\alpha_i}U_{\beta i}\right)\frac{m_i}{E}e^{-i\frac{m_i^2L}
{2E}}S,
\label{A_nu_antinu}
\end{equation}
where $E$ is the energy of the intermediate-state neutrino mass eigenstate 
which propagates a macroscopic distance $L$ and 
$S$ is an additional kinematical factor which
depends on the initial and final states. We have used the standard approximations in 
order to write the neutrino oscillation phase as a function of $m_i^2L/E$. The 
phase factor $e^{-im_i^2L/2E}$ may be thought of as the neutrino 
propagator, and, as we will see shortly, will provide the necessary CP-even phase 
discussed above.

It is important to comment that, just as in $0\nu\beta\beta$ (Eq.~(\ref{Abb})), 
the amplitude Eq.~(\ref{A_nu_antinu}) is proportional to the ``helicity suppression'' 
factor $m_i/E$. This factor reflects the fact that for either helicity of the intermediate
neutrino $\nu_i$, either the initial vertex $l_{\alpha}^+W^-\rightarrow\nu_i$, or the 
final vertex $\nu_i\rightarrow l_{\beta}^-W^+$, is helicity suppressed. In addition, a factor
like this must clearly be present in any lepton number violating process, 
given that the Majorana neutrino masses are (by assumption) the only source for lepton 
number violation, which consequently
should disappear in the limit $m_i\rightarrow 0,~\forall i$. As 
will be commented upon in more detail shortly, it is this helicity suppression factor that 
renders any observation of neutrino~$\leftrightarrow$~antineutrino oscillations 
almost impossible \cite{L}.

In order to look for explicitly CP-violating effects, we also compute the amplitude 
$\bar{A}_{\not{L}}$ for 
the CP-conjugate process $l^-_{\alpha}W^+\rightarrow l^+_{\beta}W^-$, which is given by
\begin{equation}
\bar{A}_{\not{L}}=\sum_i \left(\lambda_i U_{\alpha_i}U_{\beta i}\right)^*
\frac{m_i}{E}e^{-i\frac{m_i^2L}{2E}}\bar{S},
\label{antiA_nu_antinu}
\end{equation}
where $\bar{S}$ is identical to $S$ except, perhaps, for
an irrelevant overall phase factor. Not surprisingly,
this is very similar to the situation in $0\nu\beta\beta$, discussed in Sec.~\ref{sec_doub} 
({\it c.f.} Eqs.(\ref{Abb},\ref{anti_Abb})).

Next we restrict ourselves to the two generation case. Working in a phase convention where
the charge conjugation phase factors are trivial ($\lambda_1=\lambda_2=1$), we
parametrize the leptonic mixing
matrix $U_{\alpha i}$, $\alpha=e,\mu$, $i=1,2$, as
\begin{equation}
U=\left(\begin{array}{c c} \cos\theta& \sin\theta \\ -\sin\theta& \cos\theta\end{array}
\right)
\left(\begin{array}{c c} e^{i\frac{\xi_1}{2}}& 0 \\ 0&e^{i\frac{\xi_2}{2}}\end{array}
\right).
\end{equation}
Within this parametrization, for $\alpha=e$ and $\beta=\mu$ \cite{neutrino_antineutrino}, 
\begin{eqnarray}
A_{\not{L}}=\frac{\sin2\theta}{2}S\left[-e^{i\xi_1}e^{-im_1^2\frac{L}{2E}}\frac{m_1}{E}+
e^{i\xi_2}e^{-im_2^2\frac{L}{2E}}\frac{m_2}{E}\right], \\
\bar{A}_{\not{L}}=\frac{\sin2\theta}{2}S\left[-e^{-i\xi_1}e^{-im_1^2\frac{L}{2E}}
\frac{m_1}{E}+e^{-i\xi_2}e^{-im_2^2\frac{L}{2E}}\frac{m_2}{E}\right],
\end{eqnarray}
while the rates for the process and the anti-process are
\begin{eqnarray}
\Gamma_{\not{L}}=|A_{\not{L}}|^2=
\frac{\sin^22\theta}{4E^2}|S|^2\left[m_1^2+m_2^2-
2m_1m_2\cos\left(\frac{\Delta m^2L}{2E}-\xi\right)\right], \\
\bar{\Gamma}_{\not{L}}=|\bar{A}_{\not{L}}|^2=
\frac{\sin^22\theta}{4E^2}|S|^2\left[m_1^2+m_2^2-
2m_1m_2\cos\left(\frac{\Delta m^2L}{2E}+\xi\right)\right].
\end{eqnarray}
Here, $\xi\equiv\xi_2-\xi_1$ and $\Delta m^2\equiv m_2^2-m_1^2$. The CP-odd rate difference 
$\Delta_{CP}$ is, therefore,
\begin{equation}
\Delta_{CP}\equiv\bar{\Gamma}_{\not{L}}-\Gamma_{\not{L}}=
\frac{\sin^22\theta}{E^2}|S|^2m_1m_2
\sin\left(\frac{\Delta m^2L}{2E}\right)\sin\xi,
\label{DeltaCP}
\end{equation}
while the average rate is
\begin{equation}
\frac{\bar{\Gamma}_{\not{L}}+\Gamma_{\not{L}}}{2}=
\frac{\sin^22\theta}{4E^2}|S|^2\left[m_1^2+m_2^2-
2m_1m_2\cos\left(\frac{\Delta m^2L}{2E}\right)\cos\xi\right].
\label{anti_osc_ave}
\end{equation}

While quite simple, the results computed above contain several remarkable properties, which 
we now make explicit. First of all, answering our original question, a manifestly
CP-odd effect is present. 
As outlined in the beginning of this section, the conditions
for having such an effect are that: (i)~there must be at least two interfering
contributions to the amplitudes, (ii)~these contributions must have a CP-odd
relative phase, which here is $\xi$, that is nontrivial ({\it i.e.,} different from 
$0~\rm mod~\pi$), and (iii)~the contributions must also have a CP-even relative phase, 
which here is $\Delta m^2L/2E$, that is nontrivial. 
It is the crucial presence of this nontrivial CP-even phase, 
coming from the neutrino propagators, that allows neutrino~$\leftrightarrow$~antineutrino
oscillation to exhibit a CP-odd effect while $0\nu\beta\beta$ cannot do so.

Clearly, $\Delta_{CP}$ must vanish if either the CP-odd phase vanishes (mod~$\pi$), or the 
CP-even phase vanishes. The latter will occur if $\Delta m^2=0$ (degenerate neutrino masses) 
or if $L=0$ (vanishing travel distance).\footnote{We disregard the ``finely tuned'' points 
where $\Delta m^2L/2E=\pi,~2\pi\ldots$, which can only be chosen, in principle, for a 
monochromatic neutrino beam.} From Eq.~(\ref{DeltaCP}), we see that $\Delta_{CP}$ does indeed 
vanish when it must. Furthermore, from Eqs.~(\ref{A_nu_antinu}) and (\ref{antiA_nu_antinu}), 
we see that even in the more general case of an arbitrary number of neutrino mass eigenstates,
$\Delta_{CP}$ still vanishes, as it must, when either the CP-odd phases in the factors 
$\lambda_i U_{\alpha i}U_{\beta i}$ are equal (mod~$\pi$), or else the neutrino masses $m_i$
are all degenerate or $L=0$, so that the CP-even phases in the various terms of the amplitudes
are equal. 

Before proceeding, we pause to discuss the physical parameter space for these lepton-number
violating processes, meaning the range for the values of $\theta$, $\xi$, and the masses that 
must be probed in order to describe all the physically distinguishable values of $l_{\alpha}^+
\rightarrow l^-_{\beta}$-transitions. We will {\sl define} the mass eigenstates such that
$m_2\ge m_1$. Within this definition, the CP-odd phase-difference $\xi$ yields a potentially
different physical observable for each value in the range $[-\pi,\pi]$. It remains to discuss
what happens to the mixing angle, $\theta$. As an angular variable, it is certainly constrained
to the interval $[-\pi,\pi]$, but the general form for the amplitudes 
Eqs.~(\ref{A_nu_antinu},\ref{antiA_nu_antinu}) allow for a smaller physical range. Explicitly,
$A^{\alpha\beta}_{\not{L}}\propto\sin\theta\cos\theta$ for $\alpha\neq\beta$ or
$A^{\alpha\alpha}_{\not{L}}\propto a\sin^2\theta+b\cos^2\theta$, where $a,b\in \mathbb{C}$
(the same applies to $\bar{A}^{\alpha\beta}_{\not{L}},\bar{A}^{\alpha\alpha}_{\not{L}}$). 
The following two operations leave
physical observables ($\propto|A|^2$) unchanged: $\theta\rightarrow-\theta$ and
$\theta\rightarrow\pi-\theta$. This implies that one can choose
$\theta\in[0,\pi/2]$ and completely cover the entire physical parameter space. Note that
Eq.~(\ref{DeltaCP}) and (\ref{anti_osc_ave}) have an extra symmetry: 
$\theta\rightarrow\pi/2-\theta$, such that $\theta\in[0,\pi/4]$ yields the same results
as $\theta\in[\pi/2,\pi/4]$. This is not true, in general, for the ``diagonal'' 
$A_{\alpha\alpha}$, unless $|a|=|b|$. This situation is different from the standard
neutrino~$\leftrightarrow$~neutrino oscillations, where $\theta\in[0,\pi/4]$ fully
describes two-flavor oscillations in vacuum \cite{myreview} (as is well known, this
degeneracy is lifted if the neutrinos propagate in matter). Here, one can tell whether
the electron-type neutrino is predominantly light 
($[\theta\in[0,\pi/4]]$, the ``light side'') or heavy 
($[\theta\in[\pi/4,\pi/2]]$, the ``dark side'' \cite{darkside}) 
even if the neutrinos are propagating exclusively in vacuum. Note that one can always
choose other parametrizations, where, for example $\xi\in[0,\pi]$. The different
parametrizations are related by a ``relabeling invariance,'' which states that if one
redefines the mass eigenstates $1\leftrightarrow2$ 
([$\Delta m^2,\xi] \rightarrow [-\Delta m^2,-\xi]$) and the mixing angle 
$\theta\rightarrow\pi/2-\theta$, the amplitudes remain unchanged.

It is interesting that, as Eq.~(\ref{anti_osc_ave}) shows, lepton-number-violating 
transitions that also violate flavor ($e^{\pm}\rightarrow\mu^{\mp}$)
already occur when $L=0$. This behavior is in sharp contrast to the ordinary
lepton-number-conserving neutrino flavor oscillation. It arises from the fact that
when, for example, an incoming $e^+$ makes a neutrino, at $L=0$ the wrong-helicity
(left-handed) component of this neutrino has mass eigenstate composition proportional
to 
\begin{equation}
\sum_i \frac{m_i}{E}\lambda_iU_{ei}|\nu_i\rangle.
\end{equation}
Since the neutrino state that is a pure $|\nu_e\rangle=\sum_iU_{ei}^*|\nu_i\rangle$, it is 
clear that even at $L=0$ the left-handed component of the neutrino made by an $e^+$ is not 
a pure $\nu_e$, meaning that it contains other flavors. In the two-generation case being 
considered here, it contains a $\nu_{\mu}$ component, which can instantly produce $\mu^-W^+$.  
This fact had already been noticed by the authors of \cite{neutrino_antineutrino2}.

An especially interesting property of $\Gamma_{\not{L}}$ and $\bar{\Gamma}_{\not{L}}$
is their behavior when the neutrino mass eigenstates are degenerate: $m_1=m_2\equiv m$.
In this limit 
\begin{equation}
\bar{\Gamma}_{\not{L}}=\Gamma_{\not{L}}=
\frac{m^2\sin^22\theta}{E^2}|S|^2\sin^2\left(\frac{\xi}{2}
\right).
\label{R_short}
\end{equation} 
We see that as long as $\xi\neq 0~{\rm mod}~2\pi$, and the mixing angle $\theta\neq 0,\pi/2$,
$e^{\pm}\rightarrow\mu^{\mp}$-transitions still occur, and their rates depend on the 
mixing angle. Thus, the mixing angle continues to have physical consequences even when the
neutrino masses are degenerate.

When one remembers how quark mixing behaves, this result for Majorana neutrinos is
very surprising and puzzling. For, as one will recall, 
if the masses of all up-type quarks or down-type quarks are degenerate, mixing 
phenomena are absolutely absent. Indeed, in the presence of this degeneracy, all mixing angles 
are unphysical. This is easy to show \cite{CP_book}. 
Assume that the down-type quarks are degenerate in 
mass. This means that the relevant part of the Lagrangian is given by
\begin{equation}
{\cal L}\supset \left(-\frac{g}{\sqrt{2}} W_{\rho}^+\sum_{i,j}\bar{u}_{i L}\gamma^{\rho}
V_{ij} d_{j L} + H.c.\right) 
+ \left(-m\bar{d}_{i L}\delta_{ij}d_{j R}+H.c.\right),
\end{equation}
where $V$ is the CKM quark mixing matrix, $m$ is the hypothetical common down-type quark mass 
and $\delta_{ij}$ is the standard Kronecker-delta symbol. We can now define 
$d'_{i L}\equiv V_{ij} d_{j L}$, which renders the charged current coupling diagonal, while 
the mass term is modified to $m\bar{d'}_{i L}V_{ij}d_{j R}\equiv 
m\bar{d'}_{i L}\delta_{ij}
d'_{j R}$ if we redefine $d'_{i R}=V_{ij}d_{j R}$. 
This final redefinition does not lead
to any other physical consequences. Thus, one can choose a basis where both the down quark mass
matrix and the charged current weak couplings are diagonal.
In this basis, all mixing angles have disappeared. One may summarize the situation
in the following heuristic way: the mixing matrix $V$ may be thought of (as one
alternative) as describing the relation between the down-type quarks that have definite 
masses and those that have diagonal couplings to the up-type quarks. When the down-type
quark masses become equal, there is nothing left to describe. Any linear combination
of down-type quarks has the same mass as any other linear combination, so one
may simply choose the down-type quark basis states to be the ones whose couplings to
the up-type quarks are diagonal. There is no way to define a physically meaningful mixing
matrix.

One can try to repeat the same logic in the neutrino sector \cite{degenerate,CP_book}. 
The first step is to redefine the
neutrino states such that the charged current part of the Lagrangian Eq.~(\ref{inter}) is
diagonal: $\nu_{\alpha L}\equiv U_{\alpha i}\nu_{i L}$ (this is often referred to as
the flavor basis). In this basis, the part of the Lagrangian which contains the
Majorana neutrino mass term is given by,\footnote{This is the low-energy effective Lagrangian,
which ensues after electroweak symmetry breaking. It is independent
of the mechanism that generates Majorana neutrino masses.} 
\begin{equation}
{\cal L}\supset -\frac{1}{2}\bar{\nu^c}_{\alpha R}U^*_{\alpha i}M_{ij}U^*_{\beta j}
\nu_{\beta L}+H.c.= -\frac{1}{2}\bar{\nu^c}_{\alpha R}\left(U'^*\mathbb{E}^{-i\xi_i/2}
M\mathbb{E}^{-i\xi_i/2}U'^{\dagger}\right)_{\alpha\beta} \nu_{\beta L}+H.c.,
\label{m_alphabeta}
\end{equation}
where $U'$ and $\mathbb{E}$ have been defined in Eq.~(\ref{extract}), and $M\equiv
{\rm diagonal}(m_1,m_2,\ldots)$. If $M_{ij}=m\delta_{ij}$, no generic simplification is 
possible \cite{degenerate}. 
This means that some of the mixing angles and phases (Dirac and Majorana) which were
physical in general are still physical. The issue of counting the number of
physical parameters was discussed in detail by the authors of \cite{degenerate}, to which 
we refer readers for more details (see also \cite{CP_book}). 
One can go one step further, and ask what happens
if the model is CP invariant. This happens, for example, if 
$U'$ is real (and therefore an orthogonal matrix), and $\xi_i=0~{\rm mod}~\pi$, so that 
\begin{equation}
\left(U'^*\mathbb{E}^{-i\xi_i/2}
M\mathbb{E}^{-i\xi_i/2}U'^{\dagger}\right)_{\alpha\beta}=m
\left(U'\mathbb{E}^{i(0,\pi)}U'^{\dagger}\right)_{\alpha\beta},
\end{equation}  
where we define $\mathbb{E}^{i(0,\pi)}$ to be a diagonal matrix with elements 
equal to 1 or $-1$.\footnote{The diagonal elements of $\mathbb{E}^{i(0,\pi)}$ are also
referred to as the relative CP-parities of the different neutrino mass-eigenstates, 
which are physically meaningful \cite{CP-parities}.} 
If $\mathbb{E}^{i(0,\pi)}$ is proportional to the identity
matrix, there are no physical mixing angles. This behavior is clear in Eq.~(\ref{R_short}).
In the case $\xi=0$, such that $\mathbb{E}^{i(0,\pi)}$ is the $2\times2$~identity 
matrix, there are no lepton number violating flavor transitions
in the mass degenerate case. On the other hand,
in the CP-conserving but less trivial case of $\xi=\pi$, $\mathbb{E}^{i(0,\pi)}
\propto{\rm diagonal}(1,-1)$, and a physical mixing angle can be defined, {\it i.e.},
there are lepton number violating flavor transitions in the mass degenerate case. 

It is interesting to note that, unlike the case of degenerate Majorana neutrino masses,
if all {\sl charged lepton} masses were degenerate, there would be no physical mixing
angles (or CP-odd phases) to speak of. Indeed, if this were the case, one could always
choose a basis where the Majorana neutrino mass-matrix, the charged current coupling, and
the charged lepton mass matrix were all diagonal, 
so that $l^-_{\alpha}\rightarrow l_{\beta}^{+}$ 
transition processes would be trivially zero for $\alpha\neq\beta$. 

We summarize the situation regarding mass-degenerate neutrino eigenstates. Unlike the
situation in the quark sector, when all neutrino mass-eigenstates have the same mass,
(some of)
the mixing angles and CP-odd phases are still meaningful. A well known but under-appreciated
(at least by the authors) example of this phenomenon
is the effective neutrino mass for $0\nu\beta\beta$, 
Eq.~(\ref{Mbb}). In the three-flavor case using the standard PDG parametrization
for the mixing angles 
and assuming that the neutrino masses are degenerate\footnote{As discussed by the
authors of \cite{degenerate}, the angles and phases in the standard PDG parametrization
are not all independent if the neutrino masses are exactly degenerate.} 
(and $\lambda_i=1,~\forall i$)
\begin{equation} 
m_{\beta\beta}=m\left|\cos^2\theta_{13}\cos^2\theta_{12}
+\cos^2\theta_{13}\sin^2\theta_{12}e^{i\xi}+\sin^2\theta_{13}e^{i\zeta}\right|,
\end{equation}
where $\xi$ and $\zeta$ are the two relevant relative phases and $m$ is the common neutrino 
mass. Note that despite the degenerate masses
the two mixing angles influence $m_{\beta\beta}$, and the same 
is true of the two CP-odd phases. In the CP-preserving limit, there are several 
options. If, for example, $\xi=\zeta=0~{\rm mod}~2\pi$, $m_{\beta\beta}=m$ (no dependency on 
mixing angles), while if $\xi=0~{\rm mod}~2\pi$ and $\zeta=\pi~{\rm mod}~2\pi$, 
$m_{\beta\beta}=m|\cos2\theta_{13}|$ (dependency on (one) mixing angle). Even in the 
``effective-two-generation'' case ($\theta_{13}=0$), 
$m_{\beta\beta}$ depends on the remaining mixing angle $\theta_{12}$ so long as $\xi\neq 
0~{\rm mod}~2\pi$.

Perhaps a more intuitive way of understanding what is going on is to reinterpret the 
Majorana phases as follows. According to Eq.~(\ref{m_alphabeta}), one can rewrite
the neutrino mass matrix in the flavor basis as $U'^*M'U'^{\dagger}$, where 
$M'=M\mathbb{E}^{i\xi_i}$ is a diagonal mass matrix whose entries are in general complex.
Within this definition, the Majorana phases are interpreted as the phases of the neutrino
mass eigenvalues, which are physical if the neutrinos are Majorana particles. Within this
language, it is easy to understand what is happening in the mass-degenerate case: while
we are setting the absolute values of the masses to be equal, the mass eigenstates are
still distinguishable if the phases are different. This may even be true in the CP-conserving
case, where we can still distinguish two mass-eigenstates by the sign of the 
corresponding mass-eigenvalues. With distinguishable mass eigenstates, the mixing 
matrix still has meaning: it describes the relation between these mass eigenstates and the 
states with diagonal weak couplings to the charged leptons. This relation has physical 
consequences.

\subsection{Lepton-Number Violating Meson Decay Processes}

\begin{figure}[t]
\vspace{1.0cm}
\centerline{\epsfxsize 7cm \epsffile{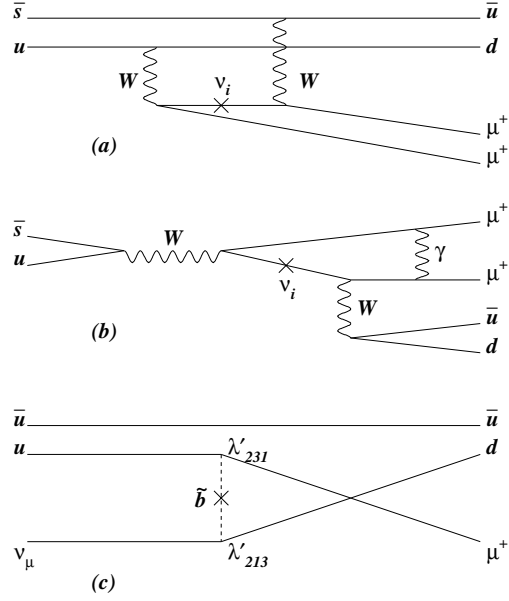}}
\caption{Feynman diagrams for $K^+\rightarrow\pi^-\mu^+\mu^+$ including
$(a)$ the leading tree-level contribution and $(b)$ a one-loop 
higher-order term (see text for details), plus (c) the Feynman diagram for 
$\nu_{\mu}\pi^0\rightarrow\mu^+\pi^-$ scattering via bottom-squark exchange,
mediated by SUSY R-parity violating $\lambda^{\prime}_{ijk}L_iQ_jD^c_k$ 
couplings. Time flows from the left to the right. The $\times$ indicates a chirality-flip
in either the neutrino or the squark propagator.}
\label{fig:meson}
\end{figure}

One can look for the effects of Majorana phases in other lepton-number violating
processes such as $K^{\pm}\rightarrow \pi^{\mp}\mu^{\pm}\mu^{\pm}$. The
present experimental upper limits on this process are at the level of
$10^{-8}$ \cite{PDG} and further improvements are likely in future. This process is
very similar to $0\nu\beta\beta$ with $e$ replaced by $\mu$.
The leading order amplitude for the $K^+$ decay is depicted in Fig.~\ref{fig:meson}a and given by 
\begin{eqnarray}
A_{\mu\mu}=\sum_i \left(\lambda_i U_{\mu i}^2\right)^* m_i K~\equiv  F e^{i\phi},
\label{Amumu}
\label{Kdecay}
\end{eqnarray}
where $K$ is a kinematical factor.
$F$ is the magnitude of $A_{\mu\mu}$, while $\phi$ is its overall CP-odd phase. This 
definition will become useful shortly.
The corresponding equation for the $K^-$ decay amplitude $\bar{A}_{\mu\mu}$ is
also given by Eq.~(\ref{Kdecay})
after replacing $(\lambda_iU^2_{\mu i})^*$ by $\lambda_iU^2_{\mu i}$ (or 
$\phi\rightarrow -\phi$). As in the case of
$0\nu\beta\beta$, since the total decay rate is given by the
absolute value of $A_{\mu\mu}$, at the leading order the CP-odd effect of
Majorana phases will be absent. However, one may consider
interference of the lowest order amplitude with the contribution of processes involving
physically accessible intermediate states, such as  
$K^+\rightarrow \mu^+ \nu$ followed by $\nu \mu^+ \rightarrow
\pi^- \mu^+\mu^+$ (depicted in Fig.\ref{fig:meson}b), or
$K^+\rightarrow \mu^+ \pi^0
\nu$ followed by $\pi^0 \nu \rightarrow \pi^- \mu^+$. 
Here, the $\Delta L=2$
interactions (and hence the Majorana phases) play a role in the second step of the
process. Let us denote this contribution by $G e^{i\psi} e^{i\gamma}$ where
the origin of $e^{i\gamma}$ (a CP-even phase factor) 
is due to the presence of the physical
intermediate state and comes from the absorptive part of the amplitude, while
$\psi$ is the CP-odd phase of this amplitude and $G$ its magnitude.

Here is the crucial point: since the intermediate state
$\mu^+\nu$ (or $\pi^0\mu^+\nu$) is a physically accessible
state, CP-even phases $\gamma$ will arise from the absorptive parts of these contributions.
This renders manifestly CP-odd effects potentially
observable, for example,  when we take the difference of the
rates for $K^+$ and $K^-$ decays. To see this explicitly, 
one can schematically write the amplitudes for the $K^{\pm}$ decays as
%\begin{eqnarray}
%|A(K^+\rightarrow \pi^-\mu^+\mu^+)|^2 = |Re A+Re X|^2 + |Im A + Im X|^2\\
%\nonumber
%|A(K^-\rightarrow \pi^+\mu^-\mu^-)|^2 =|Re A+Re X|^2 + |Im A - Im X|^2
%\end{eqnarray}
%where $A$ is the direct amplitude for $K^\pm$ decay and $X$ denotes the
%contribution from the $\mu^\pm \nu$ intermediate state. Note that the
%absorptive part does not change sign as we go from $K^+$ to $K^-$ whereas
%the direct contribution becomes complex conjugated.
\begin{eqnarray}
A(K^+\rightarrow \pi^-\mu^+\mu^+)=Fe^{i\phi}+Ge^{i\psi}e^{i\gamma}, \\
A(K^-\rightarrow \pi^+\mu^-\mu^-)=Fe^{-i\phi}+Ge^{-i\psi}e^{i\gamma}.
\end{eqnarray}
%Here, $F$ is the magnitude of the dominant amplitude (at leading order), writen 
%in Eq.~\ref{Amumu}, while $G$ is the magnitude of one amplitude which contains
%a physically accessible final state (say, the one in Fig.~\ref{fig:meson}b). $\delta$ is
%the relative CP-even rescattering phase, while $\phi$ and $\psi$ are the CP-odd
%phases (for example, $\phi$ is the phase of Eq.~(\ref{Amumu})). 
The difference $\Delta_K$ between the rates for $K^+\rightarrow \pi^-\mu^+\mu^+$ and
$K^-\rightarrow\pi^+\mu^-\mu^-$ is, therefore, 
$\Delta_K\propto 4FG\sin(\psi-\phi)\sin\gamma$  
(we have not included the phase space factors), 
as expected from the general form presented in Eq.~(\ref{Delta_gen}). 
Hence, manifestly CP-odd observables can be constructed from lepton-number violating
meson decays as long as there are relative CP-odd phases ($\phi\neq\psi$). 

As far as the generation of a CP-even phase is concerned,
the situation here is very similar to what happens in the case of leptogenesis,
where the presence of physically accessible intermediate states provides the
necessary CP-even phase which renders manifestly CP-odd effects due to 
the Majorana phases possible. We will discuss certain aspects of leptogenesis in 
the next section. 

Are there nontrivial CP-odd phases in order for the CP-odd effects to materialize?
If the only source of CP-odd phases is the leptonic mixing matrix, the
answer is, unfortunately, no: $\phi=\psi$. 
The reason is that both the ``direct decay'' (Fig.~\ref{fig:meson}a)
and the processes in which a physically accessible state is produced and later rescatters
(Fig.~\ref{fig:meson}b) have the {\sl same} CP-odd phase ($\phi$, see Eq.~(\ref{Amumu})), 
which is, therefore, unobservable. 
In order to observe a CP-odd effect, new sources of relative CP-odd phases are required.

We would like to discuss one new-physics example. Independent
$\Delta L=2$ interactions arise from supersymmetric
extensions of the SM with R-parity violation. More explicitly, a
$\lambda^{\prime}_{ijk}L_iQ_jD^c_k$ term\footnote{$L_i, Q_i$ are, respectively, lepton and
quark doublet chiral superfields, while $D^c_i$ is the down-antiquark singlet chiral superfield. 
The $\lambda^{\prime}_{ijk}$ are dimensionless couplings, and $i,j,k=1,2,3$ are family indices.} 
in the superpotential could lead
to the process $\pi^0  \nu \rightarrow \pi^- \mu^+$, depicted in Fig.~\ref{fig:meson}c,
but not directly to $K^+\rightarrow \pi^-\mu^+\mu^+$ decay, 
therefore introducing a relative CP-odd phase.
One can make an estimate of this effect using supersymmetry R-parity 
violating effects to generate the absorptive part. In this case, assuming $\psi=0$, one has
\begin{eqnarray}
\Delta_K \propto G_F^2 \sin\phi 
\frac{|\lambda^{\prime}_{231} \lambda^{\prime}_{213}| m_b}{M^3_{SUSY}},
\end{eqnarray}
where $G_F$ is the Fermi constant, $m_b$ is the mass of the bottom-quark and $M_{SUSY}$
is a supersymmetry breaking mass.
The observable effect is of course extremely tiny. Nevertheless, this example
provides a scenario where in principle CP violating Majorana phases
might lead to manifestly CP-odd effects in $\Delta L =2$ processes. Of course, one
need not restrict oneself to the kaon system, and
processes such as $B^{\pm}\rightarrow \pi^{\mp}\tau^{\pm}\tau^{\pm}$
would also work in the same way. 

Once new sources of lepton-number violation are introduced, one may
wonder whether a manifestly CP-odd observable can arise in the case of
$0\nu\beta\beta$. The answer is, in principle, yes. It would
arise, for example, from $Z\rightarrow 
(Z+2)+e^-e^-\bar{\nu}\bar{\nu}\rightarrow (Z+2)+e^-e^-$,
where the second stage is lepton number violating and is mediated by some
new kind of interaction. The CP-odd observable would be the difference between this process
and the decay of the antinucleus. It is important that the second stage be
mediated by a new form of lepton number violating interaction (so that
``$\phi \neq \psi$'') and secondly, the crucial point again
is that the intermediate state is a physical state so that we have an absorptive part.

\setcounter{equation}{0}
\setcounter{footnote}{0}
\section{Comments on Leptogenesis}
\label{sec_leptogenesis}

Perhaps the best known case of a manifestly CP-violating effect in a 
lepton-number violating process is leptogenesis \cite{leptogenesis}. 
The central idea is the following: if neutrino masses are generated via the see-saw
mechanism, there are extra singlet fermions (right-handed neutrinos) which possess
a (very heavy) Majorana mass and couple to the lepton left-handed doublet and the Higgs-boson
doublet via a Yukawa interaction. In the early Universe, these right-handed neutrinos 
will be present in the primordial thermal bath, and will eventually decay into 
leptons and scalars as soon as the Universe is cold enough. Since the decays of such states
violate lepton number, if such decays take place out of thermal equilibrium, a net lepton
number for the Universe can be generated as long as CP is also violated. Later, 
the net lepton number is converted in part to a net baryon number by nonperturbative 
sphaleron processes \cite{sphaleron}. 
For detailed reviews, we refer readers to, for example, \cite{leptoreview,leptoreview2}.

Here, we would like to concentrate on physical effects, in particular the CP-odd ones,
which are related to the 
``Majorananess'' of the processes that lead to leptogenesis. 
We would like to compare how CP is violated (and under what conditions) 
during the decay of the right-handed neutrinos to how it was violated in the 
neutrino~$\leftrightarrow$~antineutrino oscillation process analysed in Sec.~\ref{sec_osc}.
For that reason, we will concentrate on a much simpler setup, which
captures all the features we are interested in, while leaving out several unnecessary 
complications. 

We will consider the following interaction Lagrangian added to the SM one (which 
contains Eq.~(\ref{inter}) and the Yukawa coupling between the Higgs-boson, the 
lepton doublet and the right-handed charged lepton field):
\begin{equation}
{\cal L}=-\frac{M_{ij}}{2}\bar{N^c}_{iL}N_{jR}-y_{\alpha i}\left[\bar{\nu}_{\alpha L}
\bar{\varphi}^{0}-\bar{l}_{\alpha L}\varphi^-\right]N_{iR} + H.c..
\end{equation}
Here $\phi=(\varphi^+,\varphi^0)$ is the Higgs-boson weak isodoublet, $\ell_{\alpha L}=
(\nu_L,l_L)_{\alpha}$ are the left-handed
lepton doublets, and $N_{iR}$ are the right-handed neutrinos. We assume that the scalar doublet
is massless (and that $SU(2)_L$ is not broken). We will choose a basis where $M$ is diagonal,
and where its eigenvalues are real and positive. In this basis, $y$ is a generic complex 
matrix of Yukawa couplings. 

We will study the decay of the heavy right-handed neutrinos, and,
in particular, address whether CP is violated; namely, whether the branching ratio for 
$N_i \rightarrow \ell \phi$ differs from the branching ratio for $N_i \rightarrow
\bar{\ell}\bar{\phi}$, where we sum over the final state flavors of $\ell_{\alpha}$. One may 
picture the following gedanken experiment: place inside a box a certain amount of right-handed
neutrinos (of a certain ``species''). 
Wait until they all have decayed, and count the total lepton number inside the
box. If the total lepton number is not zero, CP has been violated (and lepton-number has
been ``created'').

\begin{figure}[t]
\vspace{1.0cm}
\centerline{\epsfxsize 7cm \epsffile{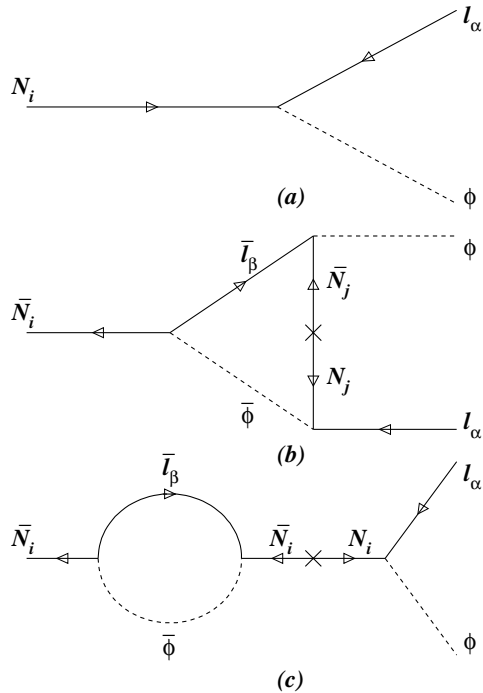}}
\caption{ Feynman diagrams for the decay of a right-handed neutrino: $(a)$ 
tree-level contribution, $(b)$ one-loop ``vertex-correction'' contribution and $(c)$ 
one-loop, ``propagator-correction'' contribution. Time flows from 
the left to the right, and the arrows represent
the chirality of the various fermions. The $\times$ indicates a chirality-flip in the
neutrino propagator, which is proportional to the neutrino Majorana mass. We only include the 
diagrams that will lead to a CP-odd contribution to the decay rate. Recall that $N_i=\bar{N}_i$
up to a phase factor.}
\label{fig:lepto}
\end{figure}

At tree-level, of course, the branching ratios are identical. At one-loop one has to compute,
on top of the tree-level contribution (Fig.~\ref{fig:lepto}a), the ``vertex-correction'' 
one-loop diagram (Fig.~\ref{fig:lepto}b) and the ``propagator-correction'' one-loop diagram 
(Fig.~\ref{fig:lepto}c). The amplitudes for $N_i\rightarrow \ell_{\alpha} \phi$ and $N_i
\rightarrow \bar{\ell}_{\alpha}
\bar{\phi}$ are, respectively, 
\begin{eqnarray}
A_{\alpha i}=\left[y_{\alpha i}+
\sum_{j,\beta}f(i,j)y_{\beta i}^*\Lambda_i^*y_{\beta j}\Lambda_jy_{\alpha j}\right]K_i, \\
\bar{A}_{\alpha i}=\left[y_{\alpha i}^*\Lambda_i^*+
\sum_{j,\beta}f(i,j)y_{\beta i}y_{\beta j}^*\Lambda_j^*y_{\alpha j}^*\right]K_i.
\end{eqnarray} 
Here, $\Lambda_i$ is the charge conjugation phase factor for the heavy Majorana mass eigenstate
$N_i$:
\begin{equation}
N_i =\Lambda_i N_i^c.
\end{equation}
The quantity $K_i$ is a kinematical factor, and $f(i,j)$ is a loop-function, which depends 
on the mass of the decaying right-handed neutrino and the mass of the
right-handed neutrino in the loop (see, for example, \cite{leptoreview,leptoreview2}). 
We have neglected terms which only serve as trivial corrections to the tree-level coupling,
and assumed the scalar and the left-handed leptons to be massless. 

The expressions for $A_{\alpha i}$ and $\bar{A}_{\alpha i}$ above may be simplified by
introducing the modified Yukawa coupling matrix $Y$, defined by
\begin{equation}
Y_{\alpha i}=y_{\alpha i}\omega_i,
\label{Y}
\end{equation}
where \cite{Nieves_Pal}
\begin{equation}
\omega_i=(\Lambda_i)^{1/2}.
\end{equation}
It is easy to show that $Y$ is unchanged by phase redefinitions of the $N_i$. Thus, rewriting
$A_{\alpha i}$ and $\bar{A}_{\alpha i}$ in terms of $Y$ makes them manifestly 
phase-redefinition invariant.

The CP-odd asymmetry in the $N_i$ decay rate, summed over all the left-handed lepton flavors
$\alpha$, is given by
\begin{eqnarray}
\Delta_{i}&\equiv&\sum_{\alpha} (|\bar{A}_{\alpha i}|^2-|A_{\alpha i}|^2)/
\sum_{\alpha} (|\bar{A}_{\alpha i}|^2+|A_{\alpha i}|^2), \\
&\simeq&\frac{2}{(Y^{\dagger}Y)_{ii}}\sum_j {\rm Im}\{(Y^{\dagger}Y)^2_{ij}\}\times 
{\rm Im}\{f(i,j)\}. 
\label{delta_i}
\end{eqnarray}
As expected, the CP-odd asymmetry is proportional to the sine of CP-odd phases
(which are present in the Yukawa coefficients $y_{\alpha i}$) {\sl and} the sine of
CP-even phases; namely, the phases of the $f(i,j)$. Do the $f(i,j)$ have complex phases? 
Fortunately, the answer is yes: the complex phase of $f(i,j)$ comes from the absorptive parts 
of both one-loop diagrams. These are indeed present since the virtual $\ell\phi$ states in 
the loop can be produced on mass-shell and rescatter.

How does this compare with CP-odd effects in neutrino~$\leftrightarrow$~antineutrino 
oscillation,
discussed in Sec.~\ref{sec_osc}? From Eq.~(\ref{delta_i}), we see that in the decay 
of the right-handed neutrino $N_i$, the interference between the tree level diagram 
Fig.~\ref{fig:lepto}a and the one-loop diagrams Figs.~\ref{fig:lepto}b and \ref{fig:lepto}c 
involving an intermediate state $N_j$ and $\ell_\beta$ leads to 
${\rm Im}[(\Lambda_iy_{\alpha i}y_{\beta i})^*(\Lambda_jy_{\alpha j}y_{\beta j})]$. 
From Eq.~(\ref{A_nu_antinu}), on the other hand, we see that in the ``oscillation'' process
$l_{\alpha}^+W^-\rightarrow l^-_{\beta}W^+$, the interference between two diagrams of the
type Fig.~\ref{diagrams}b involving, respectively, $\nu_i$ and $\nu_j$ as intermediate 
particles leads to a CP-odd contribution to the oscillation probability proportional to
${\rm Im}[(\lambda_iU_{\alpha i}U_{\beta i})^*(\lambda_jU_{\alpha j}U_{\beta j})]$.
Obviously, the imaginary parts occurring in $N_i$ decay and 
neutrino~$\leftrightarrow$~antineutrino oscillation are identical {\sl in structure}. When 
we ``transform'' from neutrino~$\leftrightarrow$~antineutrino oscillation to $N_i$ decay,
the light neutrino mixing matrix $U$, which acts as an effective coupling matrix, is simply
replaced by the heavy $N_i$ Yukawa coupling matrix $y$. Similarly, the light neutrino
charge conjugation phase factors $\lambda_i$ are replaced by the heavy $N_i$ counterparts, the
$\Lambda_i$.  

In neutrino~$\leftrightarrow$~antineutrino oscillation, each interfering diagram contains
two vertices. In $N_i$ decay, the tree-level diagram contains only one vertex, while the
one-loop diagrams with which the tree level interferes each contain three vertices. What has
happened is that in the translation between neutrino~$\leftrightarrow$~antineutrino oscillation
and $N_i$ decay, one of the vertices in one of the interfering diagrams has not only been 
replaced by its heavy $N_i$ counterpart, but has also ``left'' its original diagram, to be 
replaced by its complex conjugate in the other diagram. Of course, this ``migration'' from
one diagram to the other does not change the interference that yields the CP-odd effect.

Next, as we did in Sec.~\ref{sec_osc}, 
we examine under what conditions CP is restored. This happens, of course, in the trivial
case of only one right-handed neutrino, since $(Y^{\dagger}Y)^2_{ii}$ is real (note that the
$j=i$ term of the sum in Eq.~(\ref{delta_i}) does not contribute to $\Delta_i$). This means
that in order to violate CP, there must be nontrivial ``mixing angles'' (as is always the 
case). What do these mixing angles relate? They relate two different bases for the 
right-handed neutrino states: the mass-basis, and the ``decaying-basis,'' {\it i.e.}, the 
basis in which $Y^{\dagger}Y$ 
is diagonal.\footnote{$YY^{\dagger}$ can be chosen diagonal by redefining the $\ell_{\alpha}$
fields. This redefinition would ``resurface,'' for example, 
in the charged lepton Yukawa coupling, which does not concern us for this discussion.}

Under what conditions does $Y^{\dagger}Y$ contain no 
nontrivial phase factors? One such case is when $Y^{\dagger}Y$ is diagonal in the 
same basis where the right-handed neutrino mass is diagonal. In this case,
the decaying right-handed neutrino states coincide with the right-handed neutrino mass 
eigenstates, and one can do away with any ``mixing.'' This would happen, for example,  
if the eigenvalues $x$ of $Y$ had the same magnitude: 
$Y=V{\rm diagonal}(x,x,\ldots)U^{\dagger}$, $U,V$ unitary matrices. In this case, 
$Y^{\dagger}Y=U{\rm diagonal}(|x|^2,|x|^2,\ldots)U^{\dagger}=
{\rm diagonal}(|x|^2,|x|^2,\ldots)$. If the eigenvalues of $Y$ have the same magnitude,
one can always choose a basis where the decaying state and the mass eigenstate are the same.
Perhaps the closest analog of this situation in Sec.~\ref{sec_osc} is the case when the
charged leptons had the same mass. There, one could redefine the charged lepton states such 
that there were no $l_{\alpha}^{\pm}\rightarrow l_{\beta}^{\mp}$-transitions for 
$\alpha\neq\beta$.

Finally, we discuss the curious case of right-handed neutrinos with degenerate masses. 
Here, as in Sec.~\ref{sec_osc} when the light neutrinos all have the same mass, no 
significant simplification can be performed. In particular, similar to the situation in 
Sec.~\ref{sec_osc}, mixing angles are still generically present
%\footnote{This is true even if the mass and Yukawa 
%matrices are real. In this case, however, $\Delta_i$ would trivially vanish.} 
because of the 
presence of the Majorana phases. As we argued in Sec.~\ref{sec_osc}, one can reinterpret the 
Majorana phases as the phases of the mass-eigenvalues. In this case, even if the right-handed 
neutrino masses have the same magnitude, one can still distinguish the states by the 
phase factors. This implies that the decaying basis can still differ from the mass-basis, 
and that mixing angles may still be defined.

This means that in the case of mass-degenerate right-handed neutrinos, there is no general
reason to believe that $\Delta_i=0$. In this case, however, $\Delta_i$ (for some fixed $i$)
might not be the relevant quantity to compute if one wants to explain the excess of matter 
over antimatter in the Universe. One may, perhaps, have to compute the total lepton number
created by the simultaneous decay of all right-handed neutrinos (this is naively expected, as
they all have the same mass anyway). 

In our gedanken set-up, we proceed to analyse what happens if the ``box'' contains 
identical amounts of all the mass-eigenstate right-handed neutrinos $N_i$. 
In this case, the total lepton-number produced
by the ensuing decay of all right-handed neutrino species is (see Eq.~(\ref{delta_i}))  
\begin{equation}
\Delta\equiv\sum_i\Delta_i\propto{\rm Im}(f)\sum_{i,j}\frac{{\rm Im}\{(Y^{\dagger}Y)^2_{ij}\}}
{(Y^{\dagger}Y)_{ii}}.
\end{equation}
Here $f\equiv f(i,j), \forall i,j$ is the loop-factor, which has become independent of 
$i$ and $j$.\footnote{As was discussed by several authors, the case
of leptogenesis with mass-degenerate right-handed neutrinos is rather subtle \cite{degen_lepto,
degen_lepto2}. In particular, the 
contribution to the decay coming from the ``bubble diagram'' Fig.~\ref{fig:lepto}b is 
divergent unless one considers the decay width of the propagating right-handed neutrino. 
If the calculation is correctly performed, however, it has been shown \cite{degen_lepto2} 
that the contribution of Fig.~\ref{fig:lepto}b to $f(i,j)$ exactly vanishes in the 
mass-degenerate limit, while the vertex correction contribution Fig.~\ref{fig:lepto}c 
does not. It does satisfy $f(i,j)=f$, 
when the masses are all degenerate.} In general, $\Delta$ does not vanish, which means that,
in our gedanken set-up, a global lepton-number is generated through the decay of equal numbers
of degenerate right-handed neutrinos.

The situation here is dramatically different from the one in Sec.~\ref{sec_osc}.
There, no CP-odd effects were present in the mass-degenerate case because the CP-even phase
in the case of oscillations ($\propto \Delta m^2$) vanished exactly. Here, the CP-even 
phase does not vanish in the case of mass-degenerate states. It is curious to note, however, 
that the combination $\sum_i\Delta_i(Y^{\dagger}Y)_{ii}$ does vanish exactly when all 
right-handed neutrinos are degenerate in mass. 

Our result is independent of the number of right-handed neutrinos and left-handed leptons.
For example, the same situation occurs if instead of three right-handed neutrinos there are 
two right-handed neutrinos leading to a $3\times 2$ seesaw, which has been explored 
in several recent papers \cite{3by2} in order to try to establish a connection between
potentially measurable ``low-energy'' phases, and the CP-odd phases which are present
in the leptogenesis process. 
 
In ``thermal equilibrium leptogenesis,'' \cite{leptoreview,leptoreview2} the right-handed
neutrinos are assumed to be in thermal equilibrium with the SM fields at some
very large temperature. Under these conditions, the abundance of different 
degenerate-mass right-handed
neutrinos is guaranteed to be the same. In order to generate a net lepton number which will
later be converted to a net baryon number, the right-handed neutrinos must not only decay in 
a CP-odd fashion, but must do so out of thermal equilibrium. Do degenerate-mass right-handed
neutrinos that decay out of thermal equilibrium do so in such a way that the 
net lepton number generated is nonzero? The answer to this question is rather 
academic,\footnote{One should worry about the definition of ``degenerate'' right-handed 
neutrinos.
If the tree-level right-handed neutrino masses are all the same, quantum corrections are bound
to make them distinct, unless, for example, all right-handed neutrinos have the same 
decay-widths ({\it i.e.}, couplings). 
If this is the case, as was discussed earlier, 
there are no nontrivial mixing angles.} and beyond the intentions of this paper.

\section{Summary}
\label{theend}

In this paper we have discussed CP-violating leptonic and semileptonic
processes that can probe the CP-odd phases in the leptonic mixing matrix,
especially the so-called Majorana phases. It is nontrivial for this
probing to reveal that the Majorana phases are genuinely CP-violating
quantities. This nontriviality may be seen by looking at neutrinoless
double beta decay, which is often discussed as a way to get information on
the Majorana phases. Even though this process does depend on these phases,
in the leading order there is no difference between the rate for
the neutrinoless double beta decay of a given nucleus and that of the
corresponding antinucleus (say $^{76}$Ge
and anti-$^{76}$Ge). Therefore, these processes do not involve any {\sl
manifest} violation of CP. 

There are, however, processes which do exhibit
manifestly CP-violating effects. We have outlined the conditions under
which such effects can occur and discussed three examples: 
(i)~neutrino~$\leftrightarrow$~antineutrino oscillation, (ii)~rare leptonic decays of $K$ 
and $B$ mesons, such as $K^\pm\rightarrow \pi^\mp l^\pm l^\pm$ and
similar modes for the $B$ meson, and (iii)~leptogenesis in the early
universe, which may be responsible for the present matter-antimatter
asymmetry. 

We have discussed some limiting cases where the CP violation
disappears. A particularly interesting case, encountered in
neutrino~$\leftrightarrow$~antineutrino oscillation, is when the light neutrinos are
degenerate. We have explained why manifest CP violation is absent there. 
However, we have noted that, while the quark mixing matrix loses its
meaning when all the masses of the quarks of a given charge are of equal
size, the leptonic mixing matrix continues to have physical consequences
even when all the masses of the neutrinos are of equal size. This is
true so long as the neutrinos are Majorana particles and the relative Majorana
phases are not zero. The origin of this distinction between the behavior
of quark and lepton mixing matrices was identified. 

In the case of the CP-violation present in the decay of hypothetical right-handed neutrinos,
we also discussed under what conditions CP-violating effects would disappear. In particular,
we investigated briefly the limit of right-handed neutrinos with degenerate masses. We
comment that, in contrast to the neutrino~$\leftrightarrow$~antineutrino oscillation with
degenerate neutrino masses, CP-odd effects need not vanish. 

Admittedly, none of the ``low-energy'' processes we have considered seems to be
observable in practical laboratory experiments. However, they illustrate with 
concrete examples the important point that Majorana phases, like the more familiar 
``Dirac'' phase in the quark mixing matrix, can produce manifestly CP-violating
effects.

\section*{Acknowledgments}

We thank Leo Stodolsky for very valuable early conversations on how Majorana phases can
lead to CP-odd effects. We also thank Lincoln Wolfenstein for useful discussions. The work of
AdG and BK is supported by the US Department of Energy Contract DE-AC02-76CHO3000. 
The work of RNM is supported by the National Science Foundation Grant no. PHY-0099544.

\end{document}